\documentclass[aps,twocolumn,superscriptaddress,prl]{revtex4-2}

\usepackage{bm,graphicx,textcomp,amssymb,amsmath,dcolumn,xcolor}
\usepackage{lipsum}  
\usepackage{bbm}
\usepackage{bbold}

\usepackage{comment}
\begin{document}

\title{Pair Production in Circularly Polarized Waves}
  
\author{Christian Kohlf\"urst}

\affiliation{Helmholtz-Zentrum Dresden-Rossendorf, 
Bautzner Landstra{\ss}e 400, 01328 Dresden, Germany,}
\date{\today}

\begin{abstract}
 We study electron-positron pair production within two counterpropagating, circularly polarized electromagnetic fields through the Wigner formalism. We numerically generate high-resolution momentum maps to perform a detailed spectroscopic analysis. We identify signatures of polarization and kinematics of the incident fields in the final positron distribution and, on this basis, provide an intuitive picture of helicity transfer in multiphoton pair production. 
\end{abstract}

\maketitle
 
\paragraph{Introduction.--}

Quantum Electrodynamics (QED) might be the most accurate, and most precisely tested, theory of physics. However, despite all its success, technological limitations have left large sectors of the theory almost completely unexplored \cite{Heinzl:2009bmy, Marklund:2008gj, Gies:2008wv}. 


It was only at the turn of the millennium that first observations in non-linear QED were made in collisions of electron beams with laser light \cite{Burke, Bamber}. These experiments proved to be a milestone in the development of strong-field QED sparking renewed interest in laser-matter interactions at the extreme, see Ref.~\cite{Fedotov} for a complete overview. In the following years, further progress in technology has led to the funding of new, ultrahigh-intensity laser facilities \cite{Corels}, particle accelerators that are capable of probing purely electromagnetic interactions in ultra-peripheral collisions \cite{Star} and breakthroughs in the production of high-energy electron and photon beams \cite{Yakimenko, Luxe, XFEL}. Due to these advancements it is thus finally possible to fully explore the strong-field aspects of QED putting to a test our understanding of non-equilibrium physics, fundamental particle physics as well as emergent phenomena.  



In this Letter our focus is on multi-photon pair production, in particular, the transfer of energy, helicity and momentum from photons to the newly created electrons and positrons. \cite{footnote1}
A topic that is also of interest in atomic physics as the photon momentum transfer in the ionization of hydrogen (or positronium for that matter \cite{Moskal}) functions similarly \cite{PhysRevLett.106.193002, PhysRevLett.113.263005}. Furthermore, studying pair production is expected to generate further insight into photoelectron momentum transfer in stellar objects \cite{Seaton}, laser-plasma interactions \cite{PhysRevLett.108.165006}, quantum optics \cite{Gorlach}, and helicity transfer in electron-photon systems \cite{Keitel, PhysRevA.102.052805}.


To this end, we discuss non-perturbative pair production in high-frequent, circularly polarized fields through employing the Wigner formalism, i.e., relativistic quantum kinetic theory \cite{PhysRevLett.87.193902}. Within this approach quantum statistical as well as further collective, strong-field QED effects are already included, e.g., a field-dependent shift in the particles' effective mass \cite{PhysRevLett.109.100402}. We explicitly take into account all orders of $\hbar$ not relying on any form of gradient expansion. On the contrary, within our formalism we explicitly go beyond any dipole or even locally-constant field approximation. The particle momentum spectra, in particular, photo-positron angular distributions, are analyzed with respect to the waves' handedness and the transfer of helicity from the photon field to the particle pair. We relate the outcome of our simulations to predictions from a simple semi-classical model in order to gain a more intuitive picture of the underlying physical processes \cite{Kohlf_SpinStates}.
  
We use natural units throughout the manuscript, $\hbar=c=1$, and display results in terms of the electron mass, $m=511$ keV and the critical field strength $E_{\rm cr} = m^2/e \approx 1.3 \times 10^{18}$ V/m.

\begin{figure*}[t]
  \centering
  \begin{center}
   \includegraphics[trim={1.0cm 1cm 0.5cm 0cm},clip,width=0.99\textwidth]{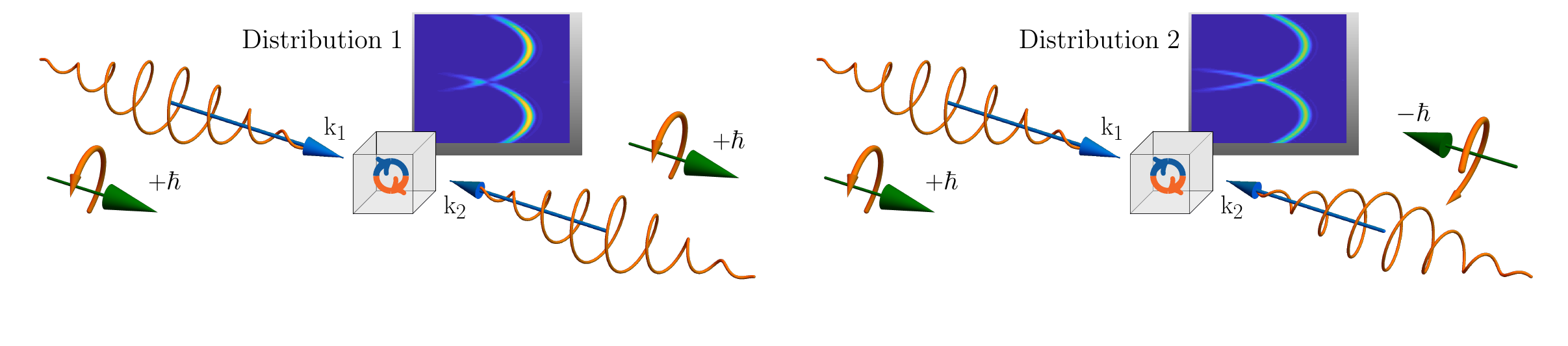} 	
  \end{center}
  \caption{Sketch of the two scenarios; {\it rl} or {\it right-handed, left-handed} on the left and {\it rr} or {\it right-handed, right-handed} on the right. Two circularly polarized fields with wave vectors ${\rm k}_1=(\omega_1, 0, 0, k_1)$ and ${\rm k}_2=(\omega_2, 0, 0, k_2)$, respectively, probe the quantum vacuum represented here as a virtual electron-positron loop. This virtual pair interacts with a discrete number of photons. Pairs that exceed the threshold for particle production become 'real observables' which can then be analyzed in terms of distribution functions. The differences in polarization of the incident waves $\pm \hbar$ manifests itself in characteristic signatures in the particles' momentum distribution, cf. the distinct color gradients in 'Distribution 1' and 'Distribution 2'. 
  \label{Sketch}
  }  
\end{figure*}
  
\paragraph{Formalism.--}

Our numerical simulations are based on evaluating the Wigner quasiprobability distribution function $\mathbbm{W} = \langle 0| \hat{\mathcal W} \left( r, p \right) | 0 \rangle$, where the quantized Wigner operator is defined as
\begin{multline}
\hat{\mathcal W}_{\alpha \beta} \left( r, p \right)  
= 
\frac{1}{2} \int {\rm d}^4 {\mathfrak s} 
\ \mathrm{e}^{\mathrm{i} p_\mu {\mathfrak s}^\mu} 
\ \mathrm{e}^{ \mathrm{i} e \int_{-1/2}^{1/2} {\rm d} 
\xi \ {\mathfrak s}^\mu A_\mu \left(r+\xi{\mathfrak s}\right) } \\
\times \left[ 
\hat{\bar\Psi}_\beta \left(r-{\mathfrak s}/2\right), 
\hat\Psi_\alpha \left(r+{\mathfrak s}/2\right) 
\right], 
\label{equ:W} 
\end{multline}
with center-of-mass $r$ and relative coordinates ${\mathfrak s}$ as well as the kinetic momenta $p$. The electromagnetic field $A_\mu$ is considered to be an external c-number field, while spinors $\hat{\bar\Psi}$, $\hat \Psi$ are quantized. The time evolution of the quantum system is determined by the Dirac equation \cite{Vasak:1987um, Birula}. Through this approach we can study fundamental pair production free of higher-order corrections, e.g., radiation or electron-electron collisions. \cite{Suppl}

 

\paragraph{Model for the Fields.--}  
 
External fields are modeled after polarized plane waves propagating in $z$-direction
\begin{equation} 
A_0 = 0,\quad {\mathbf A}(t,z) = f \left( t \right)
 \left( \begin{array}{ccc}
 \varepsilon/\omega \ \, \cos \left( \omega t + k z \right) \\ 
 \varepsilon \sigma /\omega \sin \left( \omega t + k z \right) \\
 0 \\ 
\end{array} \right) ,
\label{eq:A}
\end{equation}
where $f(t)$ is introduced as an overall damping term such that ${\mathbf A}(t,z)$ vanishes at asymptotic times guaranteeing an interpretation in terms of real particles. Periodicity confines the spatial dimension to a length $2\pi/k$. Physical parameters are the peak field strength $\varepsilon$, the field frequency $\omega = |k|$ and the quantity $\sigma$ which controls the handedness of the waves. 

A single external field of the form of Eq. \eqref{eq:A} is incapable of creating particles as both Lorentz invariants vanish. To maximize the yield, we therefore resort to configurations with two counterpropagating fields (unlike atomic ionization where often co-propagating fields are used \cite{Mancuso}). Hereby, strength and frequency of each wave are chosen independently while we consider only two configurations with respect to the fields' handedness: in the ${\it rl}-$scenario we have one right-handed and one left-handed wave, whereas the ${\it rr}-$scenario features two right-handed fields, cf. Fig. \ref{Sketch}. 

Note that the photons constituting the fields \eqref{eq:A} carry energy, momentum as well as helicity. In this regard, the proposed setup, in particular the ${\it rr}-$scenario, is much more advanced than comparable models that employ either a local or rotating dipole approximation, where pair production is evaluated at $z=0$ only \cite{Blinne, Brezin, Akal, Otto, PhysRevD.91.045016}.

\paragraph{Absorption model.--}

The Wigner formalism provides access to particle kinematics in unprecedented detail. The drawback is a reliance on simulations and an abundance of generated data which might feel overwhelming. In order to make our results more accessible, we therefore revert to an auxiliary absorption model to act as a guideline towards building an intuitive understanding. 


The basic idea is to view electron-positron pair production within external fields as an ionization of a bound state (similarly to the ionization of hydrogen or positronium). As we are only interested in a qualitative understanding at this point, 
neither the bound state's substructure nor Coulomb attraction is taken into account.


The argument is as follows, the background field has to provide enough energy for the particle to be 'ionized'. Given that there are $n$ photons that constitute a multi-photon absorption process, the total energy of these photons has to exceed the electron-positron rest mass $\sim 1.022$ MeV in order to overcome the ionization threshold. This simple relation is complicated by the fact that for high-frequency fields we have to respect the photon momentum transfer. Therefore, it is the four-vector energy-momentum conservation that has to hold \cite{Ruf, aleksandrov_prd_2016,PhysRevA.97.022515}.     
In the dipole approximation, e.g., in laser-atom ionization the momentum components are generally disregarded on the basis of different scales in the temporal and spatial domain. Only recently, the photon as a momentum carrier has drawn interest also in nonrelativistic regimes \cite{PhysRevLett.118.163203}.


 
\begin{figure}[b]
 \centering
 \begin{center}
  \includegraphics[trim={1.5cm 0.5cm 2cm 1.cm},clip,width=0.49\textwidth]{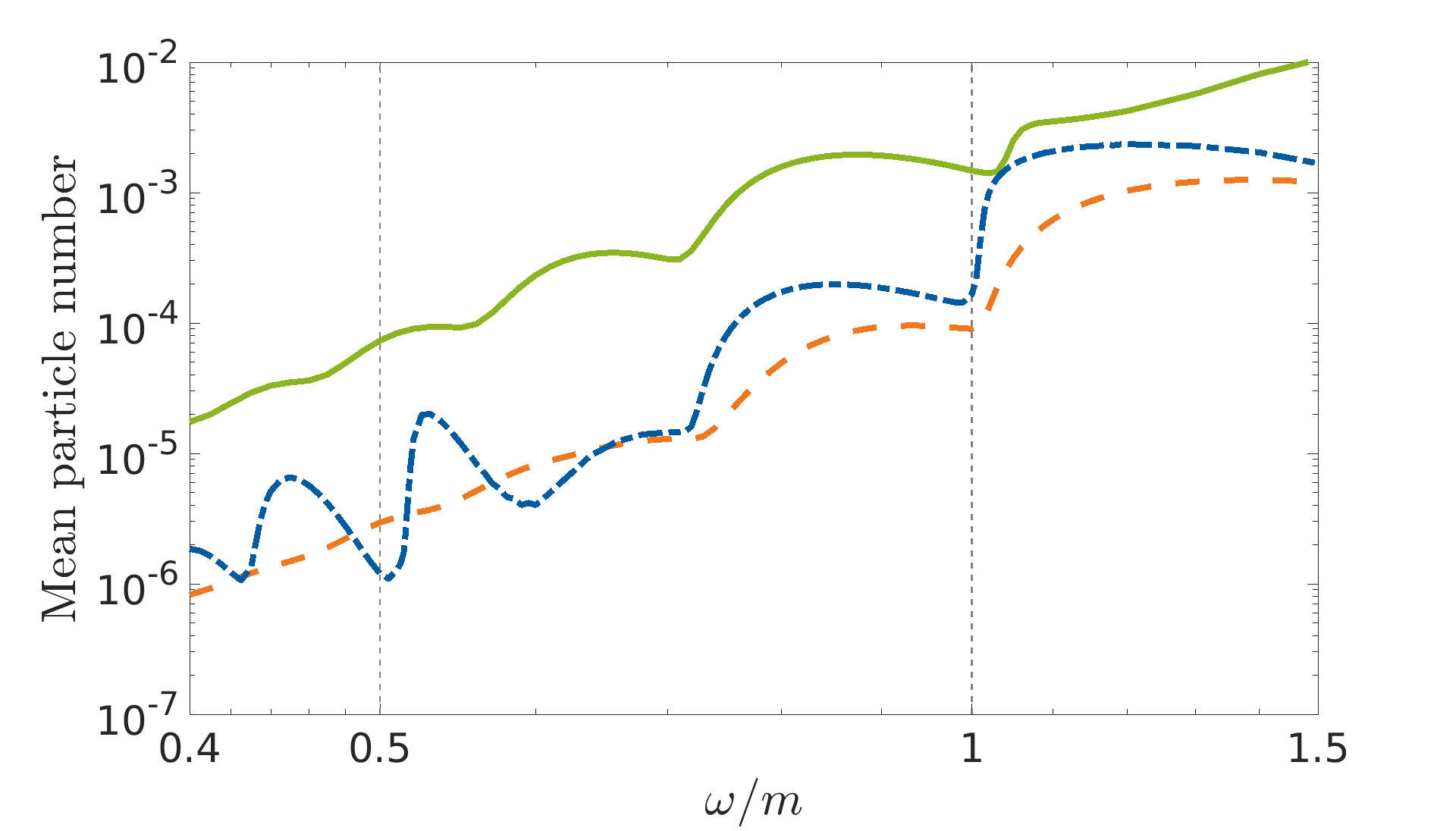} 	  
 \end{center}
 \caption{Log-log plot of the mean particle number as a function of the fields' frequencies $\omega=\omega_1=\omega_2$. The dashed, orange line gives the number of created particles in the ${\it rl}$-scenario.
 In contrast, the blue curve illustrates the result obtained for the ${\it rr}-$scenario where both waves are right-handed. 
 The green curve serves as a means for comparison displaying the normalized yield in the ${\it rl}$-scenario in a rotating dipole approximation, i.e., at the node of the standing wave where the magnetic field vanishes and the electric field is purely time-dependent. 
 Field strengths were fixed to $e\varepsilon_1=e\varepsilon_2=0.1m^2$. 
 }  
 \label{Yield}  
\end{figure} 

\begin{figure*}[t]
 \centering
 \begin{center}  
  \includegraphics[trim={1.0cm 0cm 1cm 0cm},clip,width=0.225\textwidth]{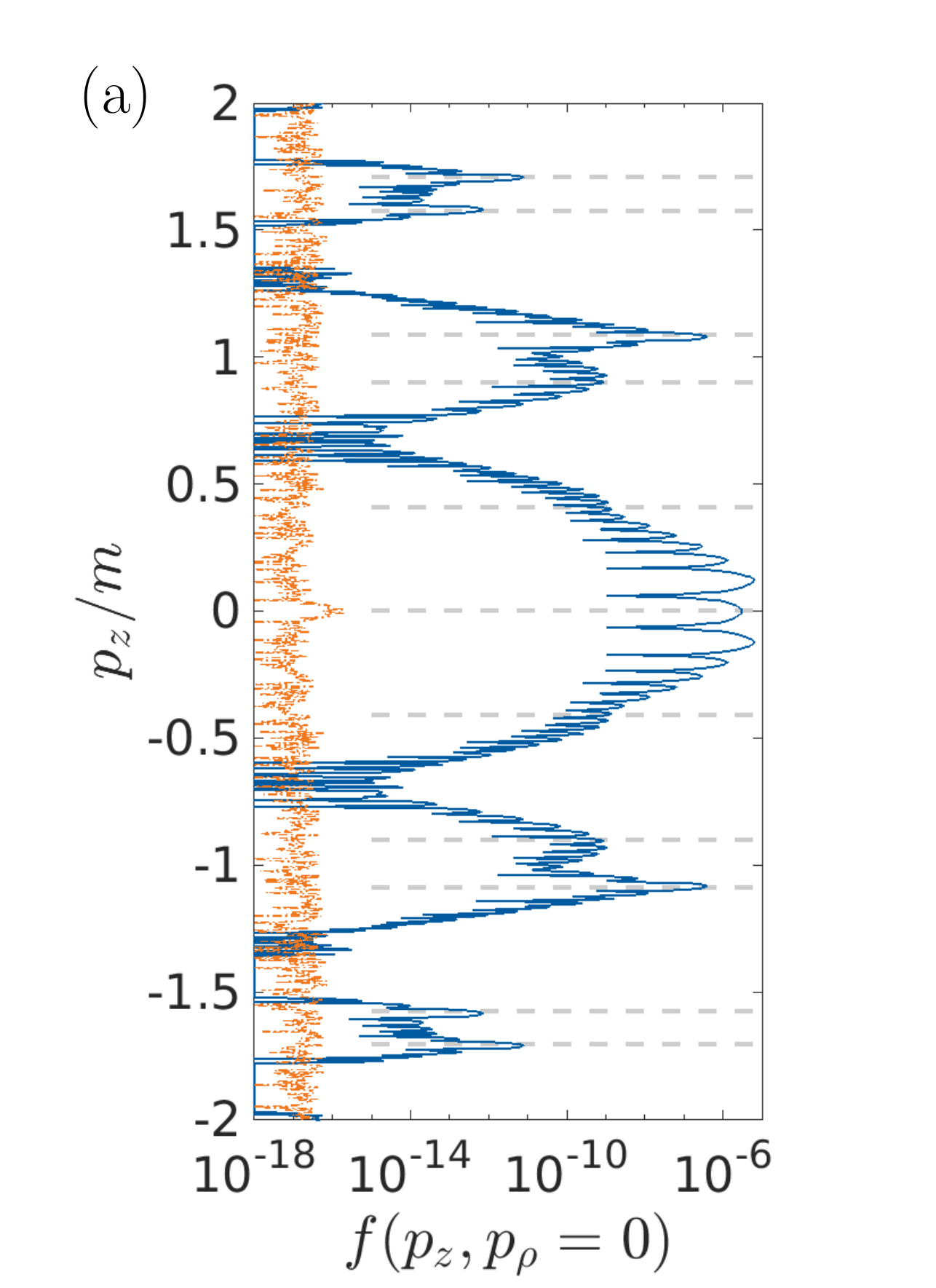} 
  \includegraphics[trim={1.15cm -0.1cm 0.55cm 0cm},clip,width=0.23\textwidth]{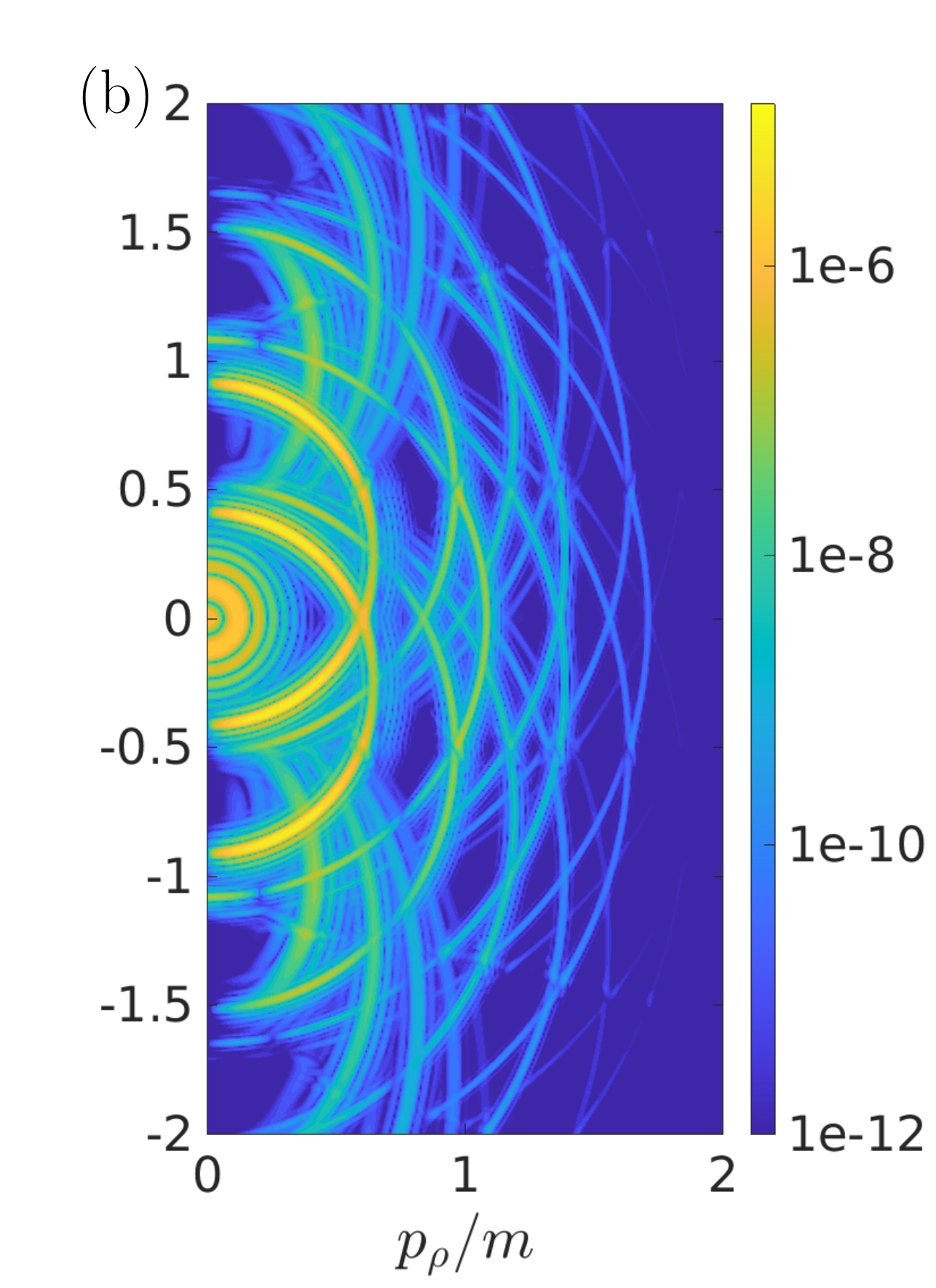} 	
  \includegraphics[trim={1.15cm -0.1cm 0.55cm 0cm},clip,width=0.23\textwidth]{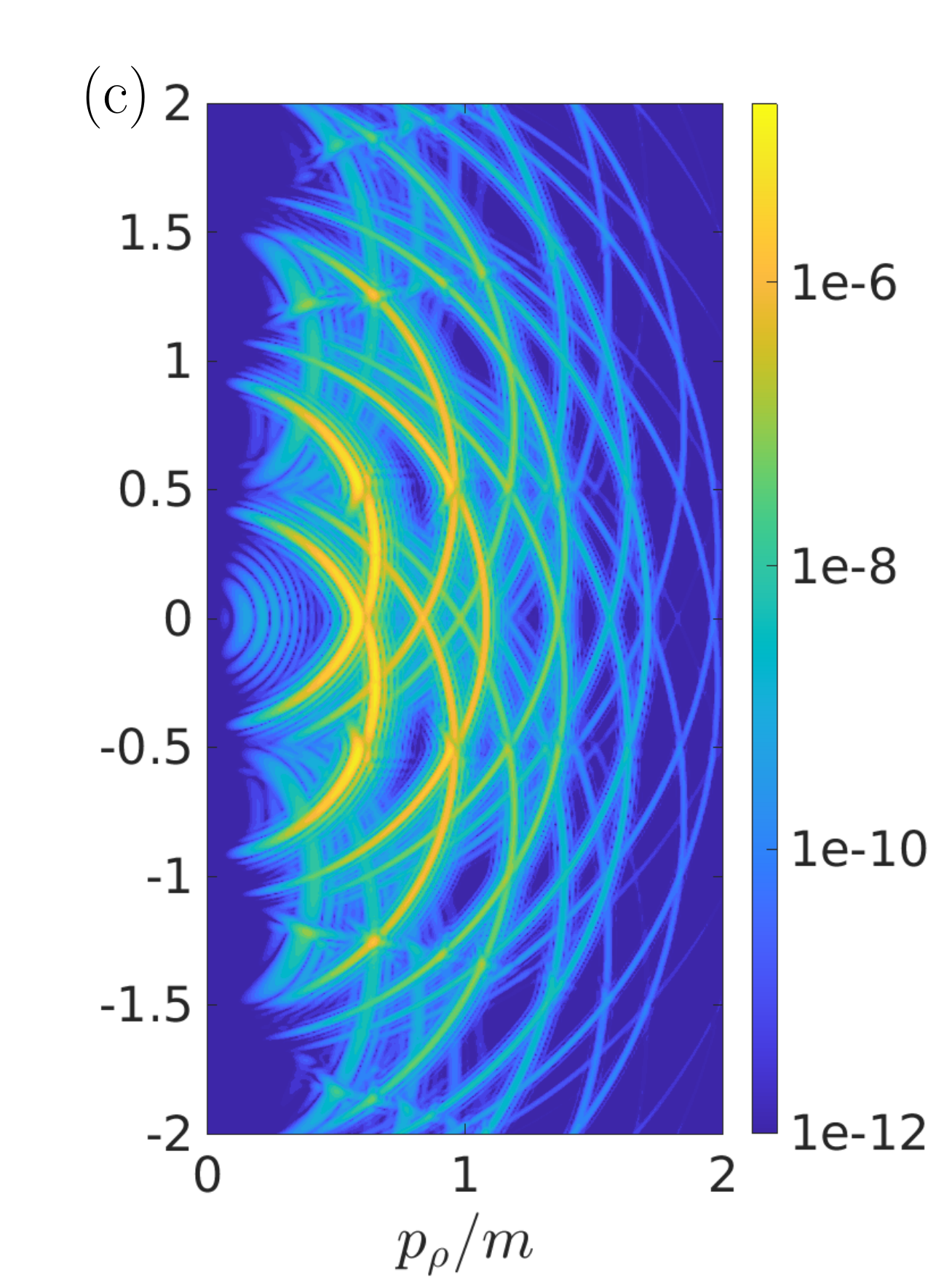} 	 	
  \includegraphics[trim={1.0cm 0.cm 1.5cm 0cm},clip,width=0.215\textwidth]{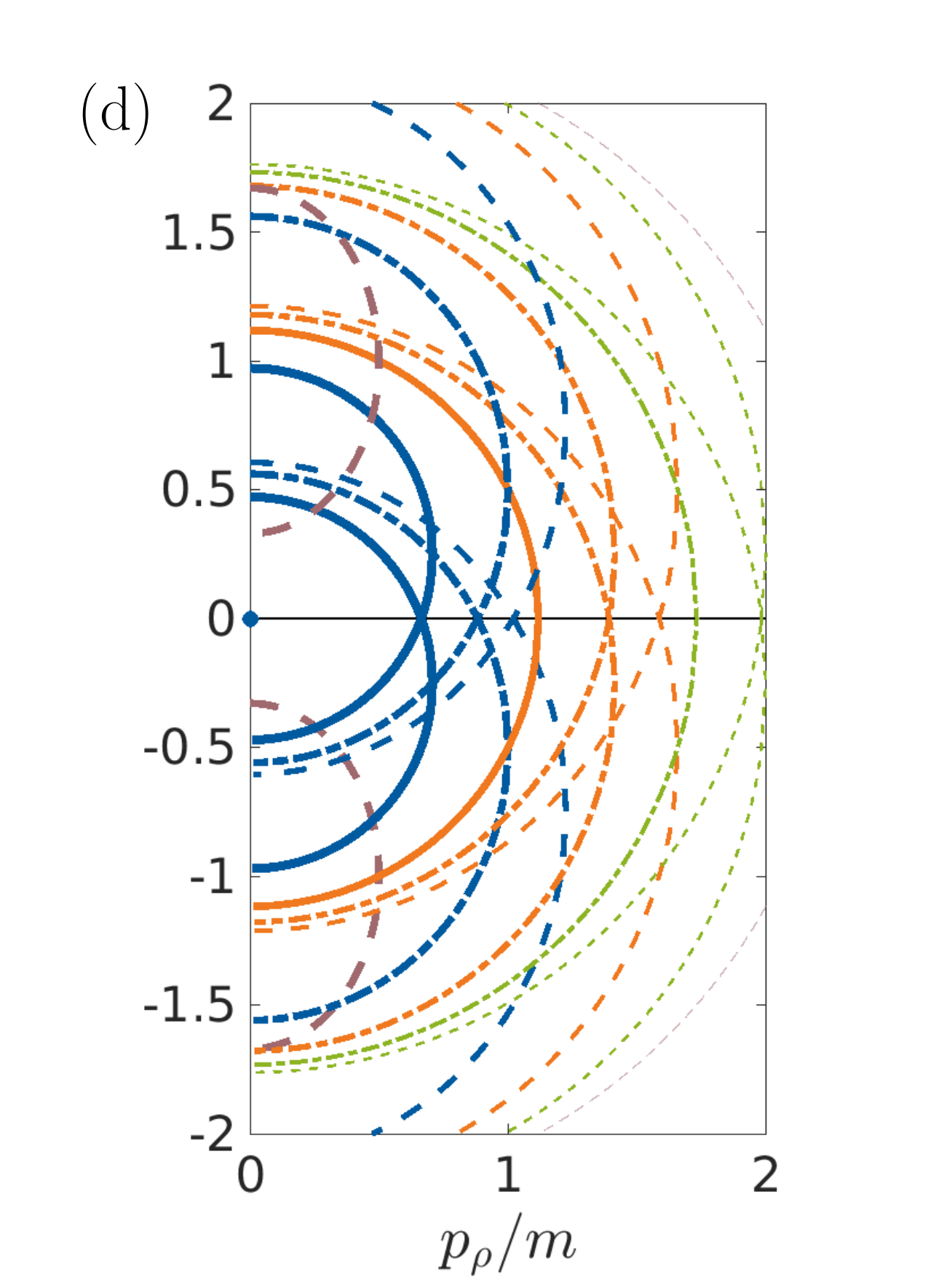} 
 \end{center} 
 \caption{Left: Particle distributions along $p_z$ for the ${\it rl}-$scenario (orange) and the ${\it rr}-$scenario (blue) with indicators for the various non-vanishing channels (channels $2-2$ at $p_z\approx0$, $2-3$ and $3-3$ at $p_z\approx m$ as well as $3-4$, $4-4$ at $p_z\approx 1.6m$).
 Middle: Particle momentum maps in $p_\rho p_z$ for the ${\it rr}$-scenario (middle left) as well as for the ${\it rl}$-scenario (middle right) further highlighting the substantial difference in particle bunching at small (transversal) momentum.  
 Right: Sketch displaying various multi-photon channels to improve orientation in the momentum maps. Structure is as follows: blue indicates channels $n-2$, orange $n-3$, green $n-4$ and brown $n-5$ (includes mirrored counterparts). The brighter the color the smaller the respective number of particles $n$.
 Field parameters: $\omega_1=\omega_2=0.5m$ and $e\varepsilon_1=e\varepsilon_2=0.1m^2$.
 }  
 \label{Map}
\end{figure*}   
 
The second important quantity is helicity or angular momentum. Photons are massless spin-1 particles, thus each photon has a helicity of $\pm 1$. The sum over which constitutes a wave's polarization. In case of a circularly polarized wave, for example, all photons exhibit the same helicity. 
Due to angular momentum conservation in multi-photon pair production this helicity is transferred to the electron-positron state where it manifests in the form of spin $S$ and orbital $L$ angular momentum. In a state with anti-parallel spin we have $S=0$, while for a state with spins aligned we obtain $S=1$; very similar to positronium \cite{Kohlf_SpinStates, PhysRevD.94.013010}. The total angular momentum is given by $J=L+S$, thus any total helicity the incident photons provide (in our scenarios $n_1 \pm n_2$), that is not used to create the intrinsic spin of the electron-positron pair, has to show up in the orbital angular momentum \cite{Reid, PhysRev.74.764}. 
  
\paragraph{Discussion.--}

It is illuminating to first study the helicity sensitivity of the total particle yield \cite{Ilchen,Krajewska}, cf. Fig. \ref{Yield}. For the sake of clarity, here we only consider symmetric configurations with both waves having a field strength $e\varepsilon=0.1m^2$ and frequency $\omega$.

In the ${\it rl}-$scenario, helicity points in direction of propagation for one wave, and opposite in the other. Thus, both waves supply the particle pair with a positive angular momentum (without restriction of generality), cf. Fig. \ref{Sketch}. This scenario is also accessible through the rotating dipole approximation, where only energy and angular momentum transfer is considered. In this regard, the latter serves as point of contact to the broader literature.
In the ${\it rr}-$scenario, in principal any integer number angular momentum can be probed. 

Overall, the mean particle number in Fig. \ref{Yield} for these two scenarios behaves similarly. There is an overall increase in the yield as a function of the field frequency, because the higher the individual photon energies the more channels are 'open' and, in turn, can contribute. The rotating dipole approximation overestimates the yield as it factors in unphysical channels, e.g., contributions from channels where either $n_1$ or $n_2$ is zero \cite{PhysRevD.101.096009}, and becomes unreliable close to the threshold $\sim 2m$ as, consequently, it allows a single photon to decay.


 
There is a big difference in the strength of the signal around an $n$-photon resonance, though. This disparity is best analyzed in terms of the particles' momentum distribution. 
Exemplary, at a frequency of $\omega=0.5m$, thus close to the $4$-photon resonance, we observe the formation of particles at rest in the $rr$-scenario, cf. Fig. \ref{Map}(b), as particles are simply unable to amount a great surplus of kinetic energy. Such a situation is hindered, however, in an $rl$-setup where instead of a clear peak at the origin the dominant contribution towards the yield comes from higher-order absorption channels, cf. Fig. \ref{Map}(c). Upon closer inspection, Fig. \ref{Map}(a), we find that the latter actually does not create any particles along the $z$-axis. 

By means of the absorption model we interpret the situation as follows. The classical orbital angular momentum is given by $\boldsymbol{L} = \boldsymbol{r} \times \boldsymbol{p}$. If a particle moves straight in $z$-direction the transversal momenta $p_\rho$ are zero, thus the criterion for a particle ejection 
along $p_z$ is that $\boldsymbol{L}$ has to vanish. 

In the $rl$-scenario, helicity is always positive hence the total angular momentum of a pair is equal the number of photons absorbed $J=n$. But as the intrinsic spin of the pair $S$ is either zero or one and due to the fact that energy-momentum conservation prevents a single photon from decaying, a pairs' orbital angular momentum is thus always non-zero $L = n - S > 0$. The necessary condition is therefore never satisfied. This situation is similar to a circularly polarized field interacting either with a quantized photon \cite{Nikishov, Villalba-Chavez:2012kko} or the field of a nucleus \cite{PhysRevA.35.4624, Yakovlev}. In both cases, the large total angular momentum impedes particle production.

\begin{figure}[t]
 \centering 
 \begin{center}
  \includegraphics[trim={0cm 0.5cm 3.5cm 2cm},clip,width=0.49\textwidth]{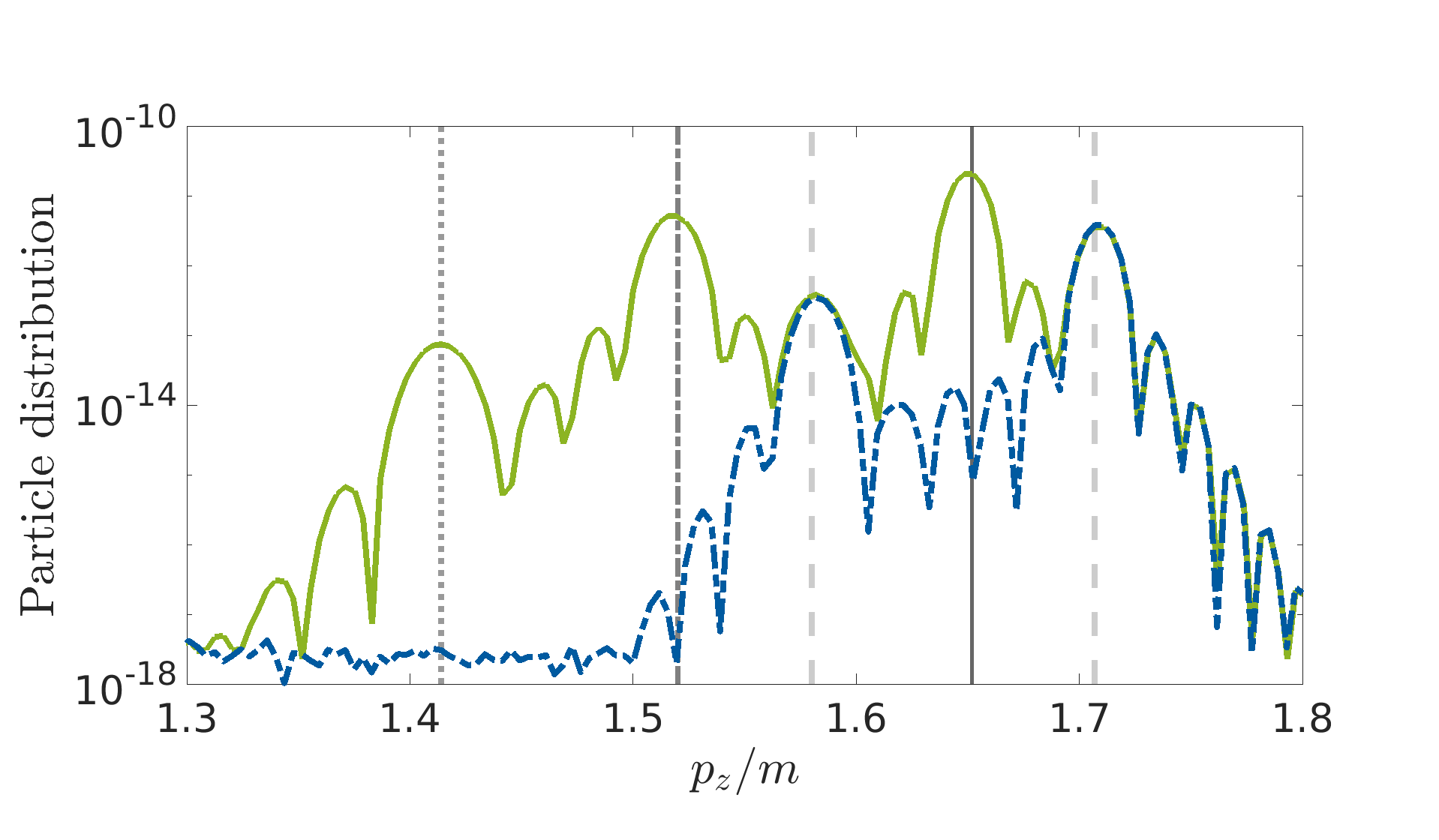} 	  
 \end{center}
 \bigskip 
 \caption{Logarithmic plot of the particle distribution for the ${\it rr}$-scenario 
 at vanishing $p_\rho$ (blue) and at $p_\rho=0.015m$ (green). The grey indicators are displayed to guide the eye, helping distinguish between spectral peaks and quantum interferences. Field parameters: $e\varepsilon_1=e\varepsilon_2=0.1m^2$ and $\omega_1=\omega_2=0.5m$.   
 }   
 \label{Channels}  
\end{figure}    


Contrariwise, in the $rr$-scenario there are multiple opportunities to form a state with $L=0$ as contributions to the angular momentum with opposite signs are indeed possible. 
In the $rr$-scenario it may be elucidating to not only study pair production along $p_z$ but perform a spectral analysis in the vicinity of $p_\rho=0$, too. In this way, channels with nonzero orbital angular momentum are revealed as well. In Fig.~\ref{Channels} such a comparison is illustrated on the basis of particle spectra around $p_z \approx 1.5m$ with $p_\rho=0$ or $p_\rho=0.015m$, respectively.

In addition to the two peaks that have already been analyzed in Fig.~\ref{Map}(a) multiple new channels appear. Data analysis suggests that these contributions stem not only from channels with angular momentum greater one. We also find evidence that states created through the same absorption channel but with a different spin $S$ feature discernible spectral peaks. This is reminiscent of other two-particle systems where a higher spin or angular momentum translates into a higher rest mass \cite{PDG}. 


\begin{figure}[t] 
 \centering
 \begin{center} 
  \includegraphics[trim={1cm 0cm 1.7cm 0cm},clip,width=0.235\textwidth]{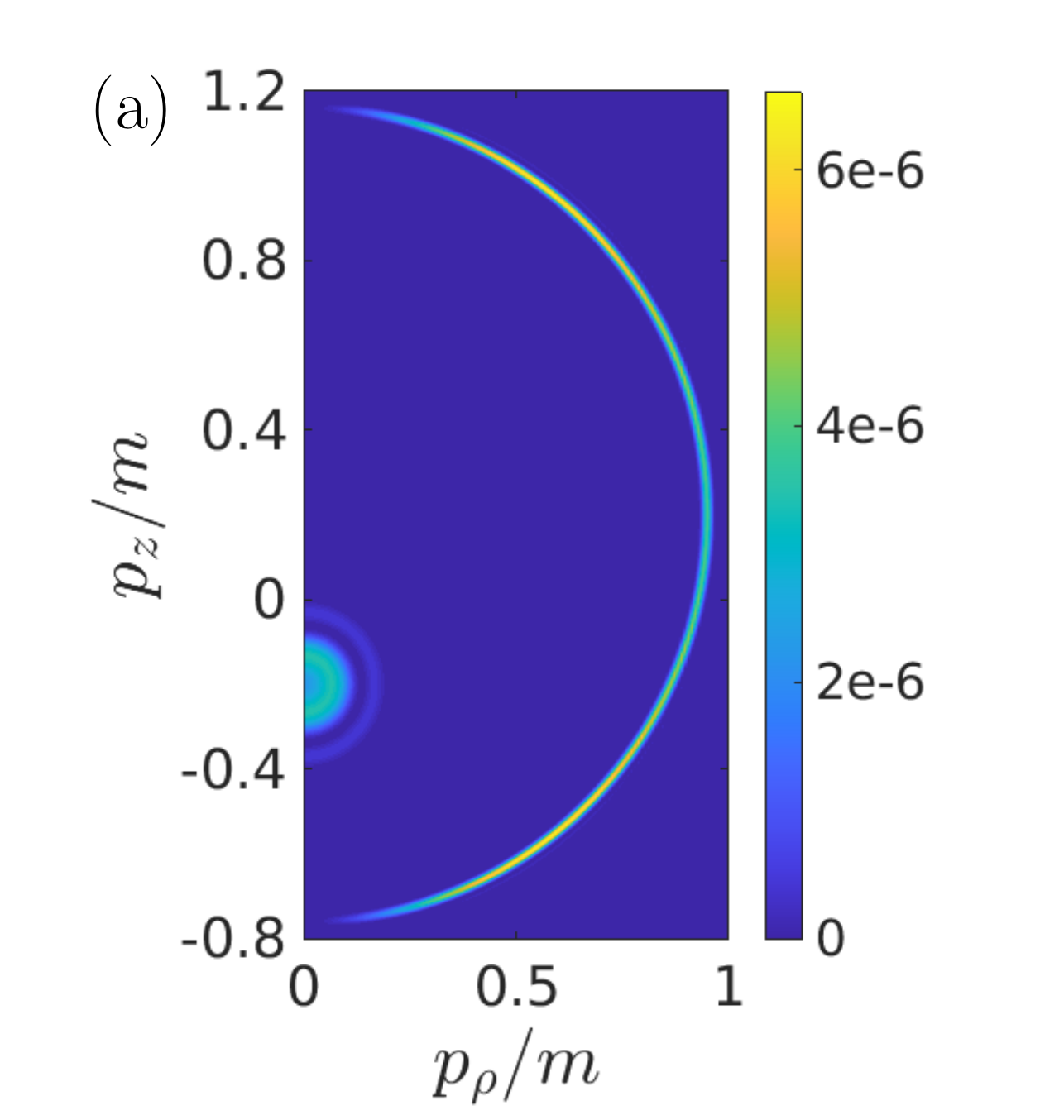} 
  \includegraphics[trim={1cm 0cm 1.7cm 0cm},clip,width=0.235\textwidth]{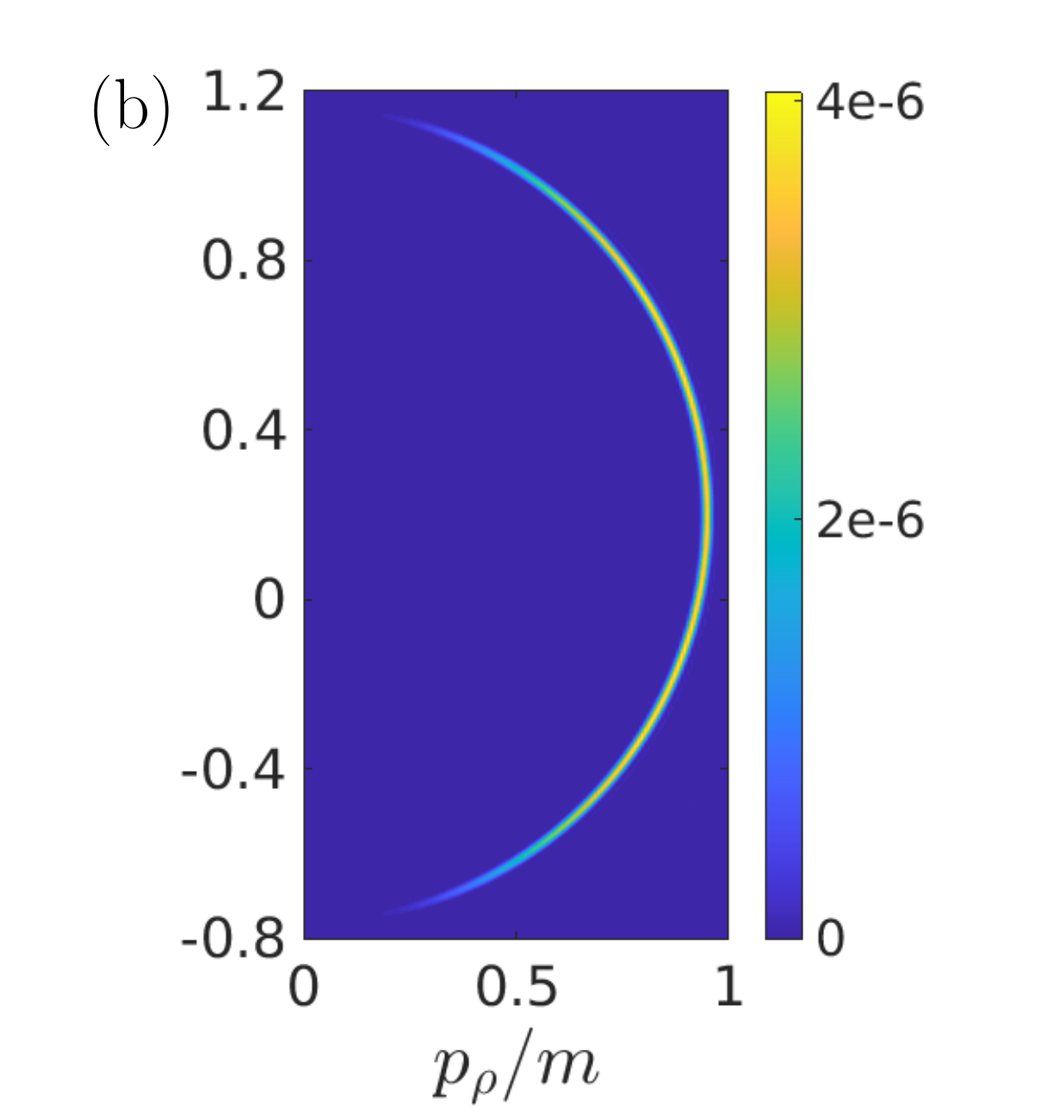}
 \end{center}
 \caption{Contour plot of positron distributions for the ${\it rr}-$scenario (left) and the ${\it rl}-$scenario (right) in a two-color set-up (peak field strengths $e\varepsilon_1=0.01m^2$, $e\varepsilon_2=0.1m^2$, frequencies $\omega_1=1.2m$, $\omega_2=0.8m$). The channel $1-2$ does not exhibit a peak at the same angle. Only in the ${\it rr}-$scenario the $1-1$ channel is contributing to the mean particle density.
 }  
 \label{Theta} 
\end{figure}  
    
Our approach is also applicable to pair creation through higher harmonics \cite{PhysRevLett.123.091802}.
For two-color configurations the positron (or electron) distributions are asymmetric in $p_z$ as $k_1 \neq k_2$. Specifically, in Fig. \ref{Theta} we have $k_1=3k_0$, $k_2=2k_0$ with fundamental frequency $k_0=0.4m$. Consequently,  particles created through the, for example, $2-1$ channel acquire a much higher kinetic energy than particles created through the $1-2$ channel. 

In the $rl$-scenario the angular momentum absorbed from the waves again stack, thus no positrons are created along $p_z$ and, in turn, no traces of the $1-1$ channel are to be found, cf. Fig. \ref{Theta}(b). 
For comparison, in the $rr$-scenario the distribution function corresponding to the $1-2$ channel peaks at a different emission angle of roughly $45^\circ$ (and $135^\circ$) but does not vanish at $p_\rho=0$ (at $p_z=1.2m$ it has retained $1\%$ of its maximal value). Additionally, $2$-photon pair production is entirely possible, cf. Fig. \ref{Theta}(a). 

This apparent difference regarding the main emission angle might also be of interest with respect to modern searches for the Breit-Wheeler process \cite{Zhao, Ribeyre, Zhu, Mercuri-Baron:2021waq, Blackburn:2021cuq}. Depending on the experimental setup it might be preferential to actively search for an optimal beam polarization such that the particle emission is maximal for a specific angle.
 
\paragraph{Conclusion.--} In this Letter, we consider pair production through two counter-propagating, circularly polarized waves. We observe that the particles' orbital angular momentum and thus their emission direction is highly sensitive to the incident waves' polarization. Ejection along the photon propagation direction is completely blocked in interactions of a right-handed with a left-handed wave. Consequently, signatures of resonances in the particle yield are smoothed and shifted towards higher frequencies, i.e., particles have to acquire a higher kinetic energy. For two right-handed waves of similar frequency no such restriction is found. This behaviour extents to two-color fields as well.
   
\acknowledgments 

\paragraph{Acknowledgments}

We want to thank Ralf Schützhold and Naser Ahmadiniaz for the interesting discussions. We are particularly grateful for the assistance given by Nina Elkina regarding High-Performance Computing. Computations have been performed on the Hemera-Cluster in Dresden. We further thank Naser Ahmadiniaz for proofreading the manuscript.
 

\end{document}